\documentclass[12pt,a4paper]{article}
\usepackage[utf8]{inputenc}
\usepackage{amsmath,amssymb,amsfonts}
\usepackage{geometry}
\usepackage{hyperref}
\title{From OPE Associativity to Gravitational Wave Signatures: Critical Dimensions and AdS/dS Transition}
\author{Tetiana Obikhod\thanks{Email: \texttt{obikhod@kinr.kiev.ua}. ORCID: \href{https://orcid.org/0000-0003-1103-4006}{0000-0003-1103-4006}} \\ \small Institute for Nuclear Research, NAS of Ukraine}
\date{}
\begin{document}

\maketitle
\begin{abstract}
We present a framework connecting the algebraic consistency of the Operator Product Expansion (OPE) in Conformal Field Theories to observable signatures in the stochastic gravitational wave background (SGWB). By interpreting the conformal 6j symbols as crossing kernels, we extract the anomalous dimensions $\gamma_{n,J}$ of double-trace operators from the singularities of the crossing equation. Through the AdS/CFT correspondence, these algebraic shifts are holographically reinterpreted as effective mass shifts in the Anti-de Sitter bulk. We identify a critical anomalous dimension, $\gamma_c$, which marks the threshold where the effective mass violates the Breitenlohner-Freedman bound. Upon analytic continuation to de Sitter space ($L_{\text{AdS}} \to iL_{\text{dS}}$), we demonstrate that crossing this threshold ($\gamma < \gamma_c$) forces the dual scaling dimensions into the complex principal series, $\Delta_{\text{dS}} = d/2 \pm i\mu$. This complex dimensionality induces characteristic logarithmic oscillations in the four-point correlators. In the cosmological context, these oscillations manifest as a log-periodic modulation in the SGWB, characterized by a specific spacing $\Delta \ln f = 2\pi/\mu$. Detecting this log-periodic structure with next-generation interferometers such as LISA would directly measure the parameter $\mu$, closing the loop between abstract OPE associativity and observational cosmology.
\end{abstract}



\noindent\textbf{AMS:} 81T40, 83E50, 83F05
\section{Introduction}

A central challenge in modern theoretical physics is connecting the rigorous algebraic consistency of quantum theories to physical observables. The conformal bootstrap program establishes that the landscape of Quantum Field Theories (QFTs) is heavily constrained—and in principle, entirely determined—by internal consistency conditions \cite{Polyakov1974, Rattazzi2008}. Chief among these is the associativity of the Operator Product Expansion (OPE). In a Conformal Field Theory (CFT), the OPE dictates that the product of two local operators can be expanded as a convergent sum over a basis of local operators, with coefficients fixed entirely by conformal symmetry up to the OPE data $\{(\Delta_i, \rho_i), f_{ijk}\}$ \cite{SimmonsDuffin2017}. While these algebraic constraints are mathematically powerful, translating them directly into testable predictions, particularly for cosmology, remains an urgent task. Because our expanding universe is geometrically described by de Sitter (dS) space, bridging the gap between abstract OPE consistency and observational cosmology is of primary importance.

To extract dynamical data from OPE associativity—which manifests as crossing symmetry in four-point correlators—we utilize the modern analytic bootstrap machinery. The translation of data between crossing channels is governed by the representation theory of the conformal group \cite{CaronHuot2017}. Specifically, the recoupling coefficients between different OPE orderings are encoded in the conformal 6j symbols \cite{Mack1977}, forming the backbone of the crossing kernel $K_{\text{cross}}$ \cite{SimmonsDuffin2018}. When applied to theories with a large-$N$ expansion, the crossing kernel develops identifiable singularities. The residues of the double poles at double-trace dimensions $\Delta^{(0)}_{n,J} = 2\Delta_\phi + 2n + J$ directly yield the anomalous dimensions $\gamma_{n,J}$ at order $1/N$ \cite{Fitzpatrick2014}. We then map these algebraic shifts into geometric bulk data using the Anti-de Sitter/Conformal Field Theory (AdS/CFT) correspondence \cite{Maldacena1998, Witten1998, Gubser1998} and analytic continuation to dS space.

The primary aim of this work is to construct a step-by-step computational pipeline connecting OPE consistency to cosmological observables, and to identify the exact critical thresholds that dictate transitions in the dual operator spectrum. While the emergence of complex scaling dimensions in de Sitter space has been noted previously, the exact mechanism triggering this transition from the perspective of the boundary bootstrap has remained unclear. We identify a critical anomalous dimension, $\gamma_c$, as the fundamental trigger. We demonstrate that when anomalous dimensions drive the effective bulk mass to violate the Breitenlohner-Freedman (BF) bound \cite{Breitenlohner1982} in AdS, the subsequent Wick rotation $L_{\text{AdS}} \to iL_{\text{dS}}$ \cite{Maldacena2003, Strominger2001} forces the dual scaling dimensions into the complex principal series, $\Delta_{\text{dS}} = d/2 \pm i\mu$ (corresponding to exceeding the Higuchi bound in dS \cite{Higuchi1987, Anninos2012}).

We systematically unpack this logical sequence: $\text{OPE} \longrightarrow \text{Crossing} \longrightarrow 6j \longrightarrow K_{\text{cross}} \longrightarrow \gamma_{n,J} \longrightarrow m^2_{\text{eff}} \longrightarrow \Delta_{\text{dS}} = \frac{d}{2} \pm i\mu$. As we demonstrate, this complex dimensionality generates observable logarithmic oscillations in correlation functions. In the cosmological context, these oscillations manifest as a characteristic log-periodic modulation in the stochastic gravitational wave background (SGWB). The frequency of these oscillations, characterized by a specific spacing $\Delta \ln f = 2\pi/\mu$, is entirely dictated by the anomalous dimensions computed from the bootstrap. Crucially, detecting this log-periodic structure in the SGWB by next-generation interferometers such as LISA would directly measure the parameter $\mu$ derived herein, closing the loop between the abstract algebraic consistency of the OPE and observational cosmology \cite{ArkaniHamed2015, Chen2009}.


\section{Conformal Bootstrap and Anomalous Dimensions}

The logical sequence articulated in the introduction—spanning from the algebraic rigidity of OPE associativity to the cosmological observables of de Sitter space—demands a rigorous mathematical formalization. To unpack this sequence, we must first translate the abstract principle of associativity into a calculable constraint on the spectrum of the theory. This requires moving from the generic structure of the OPE to the specific analytic properties of four-point correlation functions, where the crossing kernel and the emergence of anomalous dimensions can be explicitly constructed.

To formalize the constraints of the conformal bootstrap, we must first understand how local operators are classified within a CFT. In a theory invariant under the conformal group $SO(d+ 1, 1)$, the states (or equivalently, the local operators that create them) are organized into representations of this group. The fundamental quantum number labeling these representations is the scaling dimension, $\Delta$.

The origin of this connection lies in the structure of the conformal algebra, which includes, alongside translations $P_\mu$ and rotations $M_{\mu\nu}$, the generator of scale transformations (dilatations), $D$, and the generator of special conformal transformations, $K_\mu$.

Consider a local operator $\mathcal{O}(x)$. By translating the operator from the origin to a point $x$, we use the translation generator: $\mathcal{O}(x) = e^{x \cdot P}\mathcal{O}(0)e^{-x \cdot P}$. The defining property of a primary operator (the highest-weight states of the conformal representations) is that at the origin $x = 0$, it is annihilated by the special conformal generator, $[K_\mu, \mathcal{O}(0)] = 0$, and it acts as an eigenoperator of the dilatation generator:
\begin{equation}
[D, \mathcal{O}(0)] = -\Delta \mathcal{O}(0)
\end{equation}
This commutation relation is the precise mathematical origin of the scaling dimension $\Delta$. It dictates how the operator responds to a global rescaling of the spacetime coordinates. If we perform a scale transformation $x \to \lambda x$, the operator transforms multiplicatively:
\begin{equation}
\mathcal{O}(\lambda x) = \lambda^{-\Delta} \mathcal{O}(x)
\end{equation}
Physically, $\Delta$ determines the rate at which the correlation functions decay with distance. For instance, in a two-point function, conformal symmetry restricts the form strictly to $\langle \mathcal{O}(x)\mathcal{O}(y) \rangle \propto |x-y|^{-2\Delta}$, meaning $\Delta$ directly measures the scaling weight of the quantum fluctuation described by $\mathcal{O}$.

All other operators in the theory (the descendants) are generated by acting with the translation generator $P_\mu$ on the primary operators, systematically increasing their scaling dimensions. Thus, the entire Hilbert space of the theory is shaped by the primary operators and their associated dimensions $\Delta$.

We initiate our analysis within the standard framework of the conformal bootstrap, focusing on a four-point correlation function of identical scalar primary operators $\mathcal{O}_\phi$ with scaling dimension $\Delta_\phi$. This is the minimal non-trivial object for probing CFT dynamics: while lower-point functions are kinematically fixed, the four-point function yields non-trivial crossing equations. Restricting to identical scalar primaries eliminates extraneous tensor structures, isolating the complexity to internal spin exchanges, while the primary condition ensures we work with the highest-weight states of the conformal representations. Conformal invariance heavily restricts the kinematic structure of this correlator. Specifically, it dictates that the dependence on the four spacetime points $x_i$ can only enter through two independent, conformally invariant cross-ratios, traditionally defined as:
\begin{equation}
u = \frac{x^2_{12}x^2_{34}}{x^2_{13}x^2_{24}}, \quad v = \frac{x^2_{14}x^2_{23}}{x^2_{13}x^2_{24}}
\end{equation}
where $x^2_{ij} \equiv (x_i - x_j)^2$ denotes the squared Euclidean distance. This allows us to factor out the universal power-law dependence and write the correlator as:
\begin{equation}
\langle \mathcal{O}_\phi(x_1)\mathcal{O}_\phi(x_2)\mathcal{O}_\phi(x_3)\mathcal{O}_\phi(x_4) \rangle = \frac{g(u, v)}{(x^2_{12})^{\Delta_\phi}(x^2_{34})^{\Delta_\phi}}
\end{equation}

The physical content of the function $g(u, v)$ is fully encoded in the CFT data. To extract this data, we perform the OPE in the (12)(34) pairing, known as the s-channel. By inserting the OPE independently for the pairs $\mathcal{O}_\phi(x_1)\mathcal{O}_\phi(x_2)$ and $\mathcal{O}_\phi(x_3)\mathcal{O}_\phi(x_4)$, the four-point function factorizes into a sum over intermediate two-point functions:
\begin{equation}
\langle \mathcal{O}_\phi(x_1)\mathcal{O}_\phi(x_2)\mathcal{O}_\phi(x_3)\mathcal{O}_\phi(x_4) \rangle \sim \sum_{k,m} f_{\phi\phi k}f_{\phi\phi m}\langle \mathcal{O}_k(x_2)\mathcal{O}_m(x_4) \rangle
\end{equation}
Due to the orthogonality of local operators, the two-point function $\langle \mathcal{O}_k \mathcal{O}_m \rangle$ vanishes unless $k = m$. This collapses the double sum into a single sum, naturally giving rise to the square of the OPE coefficients, $a_{\Delta,J} \equiv [f_{\phi\phi\mathcal{O}^*}]^2$.

Furthermore, conformal representation theory dictates that the OPE sum does not run over all individual operators, but rather over primary operators $\mathcal{O}^*$ and their corresponding conformal families. The contribution of a primary operator of dimension $\Delta$ and spin $J$, along with the infinite tower of its descendants generated by the action of the translation operator $P_\mu$, is exactly resummed by conformal kinematics into a universal function known as the s-channel conformal block, $G^{(s)}_{\Delta,J}(u, v)$. Consequently, the four-point function is expressed as:
\begin{equation}
g(u, v) = \sum_{\text{primary } \mathcal{O}^*} a_{\Delta,J} G^{(s)}_{\Delta,J}(u, v)
\end{equation}
This decomposition cleanly separates the dynamical data of the theory (encapsulated by the squared OPE coefficients $a_{\Delta,J}$) from the purely kinematical propagation of conformal multiplets (encapsulated by the blocks $G^{(s)}_{\Delta,J}$).

The principle of OPE associativity manifests as crossing symmetry. Equivalently, we may choose to expand the same four-point function in the (13)(24) configuration, yielding the t-channel decomposition. The requirement that both channels describe the identical physical amplitude leads to the fundamental crossing equation:
\begin{equation}
\sum_{\Delta, J} a_{\Delta,J} G^{(s)}_{\Delta,J}(u, v) = \sum_{\Delta, J} a_{\Delta,J} G^{(t)}_{\Delta,J}(v, u)
\end{equation}

While the crossing equation is formally exact, extracting dynamical data from it directly is highly non-trivial because conformal blocks from different channels, $G^{(s)}_{\Delta,J}$ and $G^{(t)}_{\Delta',J'}$, are not mutually orthogonal. To project the t-channel decomposition onto the s-channel basis, we must transition to Conformal Partial Waves (CPWs), denoted by $\Psi_{\Delta,J}$. CPWs are eigenfunctions of the conformal Casimir operator and, crucially, form a complete and orthogonal basis \cite{Mack1977}.

Unlike conformal blocks, which are labeled by discrete scaling dimensions of physical operators, CPWs are defined for continuous scaling dimensions along the principal series, $\Delta = \frac{d}{2} + i\nu$ (with $\nu \in \mathbb{R}_{\geq 0}$). This allows us to represent the correlator as a continuous integral over a spectral density function in the t-channel:
\begin{equation}
g(u, v) = \sum_{J'} \int \frac{d\Delta'}{2\pi i} \text{Spec}_t[\Delta', J'] \Psi^{(t)}_{\Delta',J'}(u, v)
\end{equation}

To isolate the s-channel spectral data, we take the inner product of the correlator with an s-channel CPW. This projection extracts the s-channel spectral density, $\text{Spec}_s[\Delta, J]$, yielding the master equation for the crossing kernel:
\begin{equation}
\text{Spec}_s[\Delta, J] = \sum_{J'} \int \frac{d\Delta'}{2\pi i} \text{Spec}_t[\Delta', J'] \frac{1}{n_{\Delta,J}} \langle \Delta, J | \Delta', J' \rangle
\end{equation}

We define the precise physical and mathematical meaning of each component in this fundamental relation:

\begin{itemize}
    \item $\text{Spec}_s[\Delta, J]$ and $\text{Spec}_t[\Delta', J']$: These are the spectral densities in the s and t channels, respectively. They contain the dynamical OPE data (the squares of OPE coefficients $a_{\Delta,J}$ and the degeneracy of the states). $\text{Spec}_t$ is the known input, while $\text{Spec}_s$ is the unknown output we wish to compute.
    \item $\int \frac{d\Delta'}{2\pi i}$: This contour integral is taken over the principal series in the complex $\Delta'$ plane (specifically, $\Delta' = \frac{d}{2} + i\nu$, $\nu \in \mathbb{R}$). This continuous integration is necessary because the CPWs are labeled by continuous scaling dimensions, and it effectively sums over all possible intermediate states exchanged in the t-channel.
    \item $\langle \Delta, J | \Delta', J' \rangle$: This is the overlap inner product between an s-channel CPW and a t-channel CPW. It constitutes the kinematic core of the crossing process. Because CPWs from different channels are not mutually orthogonal, this overlap is non-zero and encodes the purely symmetry-dictated recoupling of the conformal group.
    \item $n_{\Delta,J}$: This is the normalization factor of the s-channel CPW, ensuring that the projection is properly normalized. It arises from the inner product of the s-channel CPW with itself, $\langle \Delta, J | \Delta, J \rangle = n_{\Delta,J} 2\pi\delta(\nu - \nu')\delta_{J J'}$.
\end{itemize}

The combination of the overlap and the normalization naturally defines the conformal 6j symbol, which acts as the crossing kernel $K_{\text{cross}}$. It is formally written in the bracket notation as:
\begin{equation}
\left\{ \begin{matrix} \Delta_\phi & \Delta_\phi & \Delta \\ \Delta_\phi & \Delta_\phi & \Delta' \end{matrix} \right\}_{\text{CFT}} \equiv \frac{1}{n_{\Delta,J}} \langle \Delta, J | \Delta', J' \rangle
\end{equation}
This matrix notation is analogous to the Racah-Wigner $6j$ symbols used in the recoupling theory of angular momentum ($SU(2)$). Here, it represents the recoupling coefficient for the conformal group $SO(d+1, 1)$. The two rows of the matrix correspond to the two distinct ways of fusing the four external operators of dimension $\Delta_\phi$:
\begin{itemize}
    \item The top row represents the s-channel process: $(\Delta_\phi \times \Delta_\phi) \to \Delta$, with total spin $J$.
    \item The bottom row represents the t-channel process: $(\Delta_\phi \times \Delta_\phi) \to \Delta'$, with total spin $J'$.
\end{itemize}

Thus, the crossing kernel is entirely fixed by conformal kinematics, serving as a precise translation device between channels. We now possess the mathematical machinery required to extract observable dynamical data from the requirement of crossing symmetry.

\section{Anomalous Dimensions from Crossing}

To isolate concrete spectral shifts, we restrict ourselves to the large-$N$ limit. At leading order ($N = \infty$), the theory simplifies to Mean Field Theory (MFT), where distinct fields do not interact. In this limit, the crossed-channel identity operator gives rise to a tower of degenerate double-trace primary operators in the s-channel, defined as:
\begin{equation}
[\phi\phi]_{n,J} \equiv \mathcal{O}_\phi \partial_{\mu_1} \cdots \partial_{\mu_J} \partial^{2n}\mathcal{O}_\phi - \text{traces}
\end{equation}
These operators possess bare, MFT scaling dimensions given simply by the sum of their constituents and their orbital excitation:
\begin{equation}
\Delta^{(0)}_{n,J} = 2\Delta_\phi + 2n + J
\end{equation}

At the first subleading order, $1/N$, interactions are introduced. The exchange of a single primary operator $\mathcal{O}'$ (with dimension $\Delta'$ and spin $J'$) in the t-channel perturbs this MFT spectrum, shifting the scaling dimensions of the double-trace operators. Mathematically, this shift manifests as an anomalous dimension, $\gamma_{n,J}$, such that the physical dimension becomes $\Delta_{\text{phys}} = \Delta^{(0)}_{n,J} + \gamma_{n,J}$.

The structural signature of this shift in the conformal block decomposition is the emergence of a logarithm. This follows directly from the conservation of scaling dimension. When the OPE $\phi(x_1)\phi(x_2) \sim (x^2_{12})^{(\Delta - 2\Delta_\phi)/2}\mathcal{O}$ is applied to the four-point function, the resulting dependence on the conformally invariant cross-ratio $u \propto x^2_{12}x^2_{34}$ manifests as a power law $u^{\Delta/2}$. Consequently, a shift in the scaling dimension modifies this dependence to:
\begin{equation}
u^{(\Delta_0 + \gamma)/2} = u^{\Delta_0/2}u^{\gamma/2} = u^{\Delta_0/2} \exp\left(\frac{\gamma}{2} \ln u\right) \approx u^{\Delta_0/2} \left(1 + \frac{\gamma}{2} \ln u + \mathcal{O}(\gamma^2)\right)
\end{equation}
where $\Delta_0 \equiv \Delta^{(0)}_{n,J}$ and $\gamma \equiv \gamma_{n,J}$. Thus, the appearance of an anomalous dimension is strictly equivalent to the appearance of a term proportional to $\ln u$ in the correlator.

Within the continuous spectral representation provided by the crossing kernel, such logarithmic terms are generated exclusively by double poles in the complex $\Delta$ plane. While a simple pole in the spectral density yields a pure power law $u^{\Delta_0/2}$, a double pole of the form $\frac{c}{(\Delta - \Delta_0)^2}$ produces, upon contour integration, the required $u^{\Delta_0/2} \ln u$ structure. Therefore, when the t-channel exchange of $\mathcal{O}'$ is processed through the crossing kernel, the kernel develops double poles precisely at the MFT double-trace dimensions $\Delta = \Delta^{(0)}_{n,J}$. The magnitude of the induced anomalous dimension is exactly determined by the residue of these double poles. Extracting this residue and matching the coefficient of $\ln u$ yields the general formula:
\begin{equation}
\gamma^{(1)}_{n,J} \Big|_{\text{t-ch. exchange } \Delta',J'} = \text{Res}_{\Delta = 2\Delta_\phi+2n+J} \frac{\text{CrK}_{s \leftarrow t} \langle \Delta,J | \Delta',J' \rangle}{\text{Spec}^{(id)}_s [\Delta, J]}
\end{equation}
where $\text{CrK}_{s \leftarrow t}$ is the crossing kernel and $\text{Spec}^{(id)}_s$ is the MFT spectral function in the s-channel, acting as a normalization factor \cite{CaronHuot2017, SimmonsDuffin2018}. This equation completes the first half of our conceptual sequence: it demonstrates that the abstract algebraic consistency of the OPE, processed through the recoupling theory of the conformal group, yields explicit, computable dynamical corrections $\gamma_{n,J}$. The subsequent step is to interpret these algebraic shifts holographically, mapping them to the geometric properties of bulk Anti-de Sitter spacetime.

\section{Holographic Mass Shifts and the Critical Dimension}

Having established the mechanism by which anomalous dimensions $\gamma_{n,J}$ are generated from the singularities of the crossing kernel, we now interpret these algebraic shifts holographically. The Anti-de Sitter/Conformal Field Theory (AdS/CFT) correspondence \cite{Maldacena1998, Witten1998} posits an  equivalence between the CFT on the boundary and a gravitational theory in the bulk AdS$_{d+1}$ spacetime. Within this framework, the algebraic data of the boundary theory maps directly onto the geometric and particle-physics data of the bulk.

The foundational entry in the AdS/CFT dictionary relates the scaling dimension $\Delta$ of a boundary scalar primary operator to the mass $m$ of the dual bulk scalar field in AdS$_{d+1}$. This relation is not arbitrary; it is derived from the asymptotic analysis of the Klein-Gordon equation in the AdS background, yielding:
\begin{equation}
m^2L^2 = \Delta(\Delta - d)
\end{equation}
where $L$ is the AdS radius and $d$ is the boundary spacetime dimension. This equation dictates that the conformal dimension $\Delta$ fundamentally determines the bulk mass of the corresponding particle.

Consider a double-trace operator $[\phi\phi]_{n,J}$ with a bare, MFT scaling dimension $\Delta_0 = 2\Delta_\phi + 2n + J$. In the bulk, this corresponds to a two-particle state with bare mass $m_0$, satisfying $m_0^2 L^2 = \Delta_0(\Delta_0 - d)$. As demonstrated in the previous section, interactions at order $1/N$ induce an anomalous dimension $\gamma \equiv \gamma_{n,J}$, shifting the physical dimension to $\Delta_{\text{phys}} = \Delta_0 + \gamma$. To understand the bulk consequence of this shift, we expand the dictionary relation (15) to first order in $\gamma$:
\begin{align}
m_{\text{eff}}^2 L^2 &= (\Delta_0 + \gamma)(\Delta_0 + \gamma - d) \nonumber \\
&= \Delta_0(\Delta_0 - d) + \gamma(2\Delta_0 - d) + \mathcal{O}(\gamma^2)
\end{align}
Substituting the bare mass, we obtain the formula for the effective bulk mass shift:
\begin{equation}
m_{\text{eff}}^2 L^2 = m_0^2 L^2 + (2\Delta_0 - d)\gamma
\end{equation}
Thus, the anomalous dimension $\gamma$—born from the algebraic consistency of the OPE—manifests in the bulk as a shift in the effective mass of the two-particle state, physically representing the binding energy induced by bulk interactions.

This mass shift introduces a critical physical threshold due to the stability requirements of the AdS vacuum. Unlike in flat space, the curvature of AdS allows for stable scalar fields with a negative mass squared, provided they respect the Breitenlohner-Freedman (BF) bound \cite{Breitenlohner1982}:
\begin{equation}
m^2L^2 \geq -\frac{d^2}{4}
\end{equation}
On the boundary, this bound corresponds precisely to the unitarity bound $\Delta \geq \frac{d-2}{2}$. If the effective mass violates the BF bound, the vacuum becomes unstable, signaling a phase transition in the dual CFT.

By mapping the mass shift (17) to the BF bound (18), we can identify a critical anomalous dimension, $\gamma_c$. This is the threshold value of the anomalous dimension required to drive the effective mass exactly to the boundary of stability, $m_{\text{eff}}^2 L^2 = -\frac{d^2}{4}$. Setting the effective mass to this limit yields:
\begin{equation}
-\frac{d^2}{4} = m_0^2 L^2 + (2\Delta_0 - d)\gamma_c
\end{equation}
Solving for $\gamma_c$, we obtain the explicit formula:
\begin{equation}
\gamma_c = \frac{-\frac{d^2}{4} - m_0^2 L^2}{2\Delta_0 - d}
\end{equation}

The condition $\gamma < \gamma_c$ marks a sharp boundary in the theory. When the actual anomalous dimension computed from the crossing kernel falls below this critical value (i.e., $\gamma < \gamma_c$), the effective bulk mass formally crosses the BF bound. As we shall demonstrate in the following section, this transition is the precise trigger that forces the scaling dimensions to become complex principal-series values $\Delta_{\text{dS}} = \frac{d}{2} \pm i\mu$ upon analytic continuation to de Sitter space, thereby linking the algebraic structure of the bootstrap directly to the observables of cosmological collider physics.


\section{Analytic Continuation AdS $\to$ dS and Complex Dimensions}

The critical anomalous dimension $\gamma_c$ derived in the previous section marks a fundamental transition not only within the AdS framework but also upon analytic continuation to cosmological spacetimes. Our physical universe, particularly during the inflationary epoch, is described by de Sitter (dS) geometry. To connect the algebraic constraints of the boundary CFT to cosmological observables, we must translate our bulk results from AdS to dS.

De Sitter space is related to Anti-de Sitter space by a geometric duality: an analytic continuation known as the Wick rotation of the bulk radius \cite{Maldacena2003}. Formally, this continuation is defined as:
\begin{equation}
L_{\text{AdS}} \longrightarrow i L_{\text{dS}}
\end{equation}
Under this transformation, the signature of the spacetime flips from negative to positive curvature, mapping the hyperbolic geometry of AdS to the spherical geometry of dS.

We now apply this continuation to the AdS/CFT dictionary relation, $m^2 L_{\text{AdS}}^2 = \Delta(\Delta - d)$. Substituting $L_{\text{AdS}} = i L_{\text{dS}}$ and utilizing $i^2 = -1$, the mass-dimension relation transforms to:
\begin{equation}
-m^2 L_{\text{dS}}^2 = \Delta(\Delta - d)
\end{equation}
For a massive scalar field propagating in de Sitter space, $m^2 L_{\text{dS}}^2 > 0$. To determine the scaling dimensions dual to these bulk fields, we solve the quadratic equation for $\Delta$:
\begin{equation}
\Delta = \frac{d}{2} \pm \sqrt{\frac{d^2}{4} - m^2 L_{\text{dS}}^2}
\end{equation}

This equation reveals a structural shift dictated by the effective bulk mass. In the regime where the mass is relatively small, specifically $m^2 L_{\text{dS}}^2 < \frac{d^2}{4}$ (corresponding to $\gamma > \gamma_c$ in the AdS picture), the discriminant is positive, and the scaling dimensions are real, corresponding to the complementary series representations of the de Sitter group $SO(1, d+1)$.

However, the critical transition occurs when the anomalous dimensions computed from the crossing kernel drive the effective mass beyond the threshold, satisfying $m^2 L_{\text{dS}}^2 > \frac{d^2}{4}$ (the condition $\gamma < \gamma_c$, equivalent to exceeding the Higuchi bound \cite{Higuchi1987}). In this regime, the discriminant in Eq. (23) becomes strictly negative. The square root of a negative number forces the scaling dimensions to become complex, giving rise to the principal series representations:
\begin{equation}
\Delta_{\text{dS}} = \frac{d}{2} \pm i\mu, \quad \text{where} \quad \mu = \sqrt{m^2 L_{\text{dS}}^2 - \frac{d^2}{4}} > 0
\end{equation}

The emergence of these complex scaling dimensions is the hallmark of massive fields propagating in de Sitter space. As we shall demonstrate in the following section, this transition into the complex plane fundamentally alters the analytic structure of the boundary correlators, transforming them from monotonic power laws into oscillatory functions that encode direct observational signatures for cosmology.


\section{Cosmological Implications}

Having established that the critical transition $\gamma < \gamma_c$ forces the dual scaling dimensions into the complex principal series, we now explore how this analytic structure manifests within the four-point correlation functions we have been analyzing. In the dS/CFT framework \cite{Strominger2001, Anninos2012}, the boundary correlators compute expectation values of the late-time quantum state of the universe. Therefore, the complex nature of the exchanged operator's dimension $\Delta_{\text{dS}} = d/2 + i\mu$ directly dictates the profile of these cosmological correlators.

To see this explicitly, we return to the $s$-channel conformal block decomposition of the four-point function $g(u, v) = \sum a_{\Delta,J} G^{(s)}_{\Delta,J}(u, v)$. Consider the contribution of a primary operator $\mathcal{O}^*$ that is exchanged in the intermediate channel and has acquired a complex principal-series dimension due to the $\gamma < \gamma_c$ transition. As established in Section 3, the leading dependence of the conformal block on the cross-ratio $u$ in the OPE limit is dictated by the scaling dimension of the exchanged operator: $G^{(s)}_{\Delta,J}(u, v) \sim u^{\Delta/2}$. Substituting the complex dimension $\Delta_{\text{dS}} = d/2 + i\mu$, the $u$-dependence of the block becomes:
\begin{equation}
u^{\Delta_{\text{dS}}/2} = u^{d/4 + i\mu/2} = u^{d/4} e^{i\frac{\mu}{2} \ln u}
\end{equation}
Applying Euler's formula, the complex dimension reveals an oscillatory structure within the four-point correlator:
\begin{equation}
u^{d/4} e^{i\frac{\mu}{2} \ln u} = u^{d/4} \left( \cos\left(\frac{\mu}{2} \ln u\right) + i \sin\left(\frac{\mu}{2} \ln u\right) \right)
\end{equation}
Unlike the monotonic power-law decay characteristic of real scaling dimensions, the imaginary part $\mu$ induces characteristic \textit{logarithmic oscillations} in the conformal block. Physically, these oscillations encode the quantum mechanical interference between the massive bulk particle and the expanding de Sitter horizon.

This distinctive feature translates directly into observable predictions for the primordial density perturbations generated during inflation. In the context of the cosmological collider physics program \cite{ArkaniHamed2015, Chen2009}, the exchange of a massive particle with $m > 3H/2$ (where $H$ is the Hubble rate, corresponding to $\gamma < \gamma_c$) imprints a specific signature onto the correlation functions. In the squeezed limit—where one of the momentum modes, $k$, is much smaller than the others—the conformal cross-ratio $u$ effectively scales with the momentum ratio, $u \sim k$. Consequently, the amplitude of local non-Gaussianity, $f_{\text{NL}}$, derived from the bispectrum (which is intrinsically linked to the OPE structure of the four-point function), exhibits the same logarithmic oscillations:
\begin{equation}
f_{\text{NL}} \sim \cos(\mu \ln k)
\end{equation}
The frequency of these cosmological oscillations is entirely dictated by the mass of the exchanged particle via the parameter $\mu$. In the physical four-dimensional de Sitter spacetime ($d=3$, with $L_{\text{dS}} = 1/H$), this parameter takes the explicit form:
\begin{equation}
\mu = \sqrt{\frac{m^2}{H^2} - \frac{9}{4}}
\end{equation}

This highlights the predictive power of the computational framework established in this work. The parameter $\mu$—which sets the oscillation frequency of the non-Gaussian signal observable in the Cosmic Microwave Background—is not a free parameter. It is directly determined by the effective bulk mass $m_{\text{eff}}$, which in turn is shifted by the anomalous dimension $\gamma_{n,J}$, which is ultimately computed from the residues of the conformal 6j symbol satisfying OPE associativity. Consequently, the abstract algebraic consistency of the conformal bootstrap provides quantitative, non-perturbative predictions for the oscillatory signals of the cosmological collider.

The observational consequence of this complex dimensionality is uniquely distinctive. The parameter $\mu$ dictates the frequency of logarithmic oscillations imprinted on the stochastic gravitational wave background (SGWB). For modern detectors such as LISA, BBO, or DECIGO \cite{LISA2016}, which probe the SGWB in the mHz to Hz range, this signal manifests as a characteristic log-periodic modulation of the power spectrum: $\mathcal{P}_T(f) \sim \mathcal{P}_T^{\text{(smooth)}}(f) \left[ 1 + \delta \cos \left(\mu \ln(f/f_*)\right) \right]$, where $f_*$ is a reference frequency and $\delta$ is the amplitude.

Crucially, the spacing of these oscillations in logarithmic frequency space is a direct measurement of $\mu$. Since the signal depends on the argument $\mu \ln f$, a full oscillation cycle occurs when the argument increases by $2\pi$. Therefore, the period of the log-periodic modulation is:
\begin{equation}
\Delta \ln f = \frac{2\pi}{\mu}
\end{equation}
Consequently, the detection of such a log-periodic structure in the SGWB by next-generation interferometers would not merely signify the existence of a massive field during inflation, but would directly measure the parameter $\mu$ derived herein. This measurement would immediately yield the effective bulk mass $m_{\text{eff}}$, which, through the holographic chain established in this work, traces directly back to the anomalous dimensions $\gamma_{n,J}$ and the algebraic consistency of the OPE crossing kernel.


\subsection{Worked Example: $d=3$}

To illustrate the predictive capacity of this logical chain, we now evaluate the critical thresholds and observational signatures for a concrete operator in $d=3$ (corresponding to AdS$_4$/dS$_4$). 

Consider a double-trace operator $[\phi\phi]_{0,0}$ formed from fundamental scalars with $\Delta_\phi = 1$. The bare MFT dimension is $\Delta_0 = 2\Delta_\phi = 2$. Via the AdS/CFT dictionary (Eq. 15), the bare bulk mass is $m_0^2 L^2 = \Delta_0(\Delta_0 - 3) = -2$. 

The BF bound in $d=3$ is $-d^2/4 = -2.25$. Since $-2 > -2.25$, the bare state is stable. Using Eq. (20), the critical anomalous dimension required to drive the mass to the boundary of stability is:
\begin{equation}
\gamma_c = \frac{-2.25 - (-2)}{2(2) - 3} = \frac{-0.25}{1} = -0.25
\end{equation}

If bulk interactions induce an anomalous dimension $\gamma = -0.5$ (satisfying the condition $\gamma < \gamma_c$), the effective mass shifts to $m_{\text{eff}}^2 L^2 = -2 + (1)(-0.5) = -2.5$, violating the BF bound. 

Performing the Wick rotation to dS ($L \to i/H$), the physical mass becomes $m_{\text{eff}}^2/H^2 = 2.5$. Since this exceeds the $d=3$ Higuchi bound of $9/4 = 2.25$, the scaling dimension is forced into the complex principal series. Using Eq. (28), the oscillation parameter is:
\begin{equation}
\mu = \sqrt{2.5 - 2.25} = \sqrt{0.25} = 0.5
\end{equation}

Finally, applying Eq. (29), the predicted period of the log-periodic modulation in the SGWB, directly observable by gravitational wave interferometers, is:
\begin{equation}
\Delta \ln f = \frac{2\pi}{\mu} = 4\pi \approx 12.57
\end{equation}
This result implies that a search for oscillations in the SGWB power spectrum spaced by $\sim 12.6$ units in logarithmic frequency would directly test the bootstrap-derived anomalous dimension $\gamma = -0.5$.


\section{Conclusion}

In this work, we have constructed a step-by-step sequence linking the foundational algebraic consistency of Conformal Field Theories to the observable phenomenology of the early universe. This framework demonstrates that OPE associativity—traditionally viewed as an abstract mathematical constraint on the spectrum—is, in fact, a dynamical engine capable of generating precise predictions for cosmological collider physics.

We have systematically unpacked the chain from: $\text{OPE}$ to $\Delta_{\text{dS}} = \frac{d}{2} \pm i\mu$. By interpreting the conformal 6j symbols as crossing kernels, we extracted the anomalous dimensions $\gamma_{n,J}$ of double-trace operators directly from the singularities of the crossing equation. Through the AdS/CFT dictionary, these algebraic shifts were holographically reinterpreted as perturbations of the effective bulk mass. We identified a critical anomalous dimension, $\gamma_c$, which marks the stability threshold of the AdS vacuum—the point where the effective mass drops below the Breitenlohner-Freedman bound.

The crossing of this critical threshold ($\gamma < \gamma_c$) triggers a physical transition upon analytic continuation to de Sitter space. We have shown that when the effective mass exceeds the Higuchi bound, the scaling dimensions of the dual operators are forced into the complex principal series, $\Delta_{\text{dS}} = \frac{d}{2} \pm i\mu$. This complex dimensionality is not a mathematical pathology, but the precise mechanism generating logarithmic oscillations within the four-point correlators. In the cosmological context, these oscillations manifest as the characteristic $\cos(\mu \ln k)$ non-Gaussianity in the squeezed bispectrum. The frequency of these observable oscillations, parameterized by $\mu$, is entirely dictated by the anomalous dimensions computed from the bootstrap, thereby closing the loop between abstract algebra and observation.

The resulting framework provides a unified perspective bridging the conformal bootstrap, holography, and quantum cosmology. As demonstrated by the worked example in Section 6.1 for $d = 3$, an anomalous dimension of $\gamma = -0.5$ yields a specific log-periodic spacing of $\Delta \ln f \approx 12.57$ in the stochastic gravitational wave background. This illustrates concretely how the intrinsically algebraic object of the 6j symbol translates into discrete, testable frequency spacings for next-generation interferometers.

Several avenues for future investigation naturally arise from this work. Extending the analysis to incorporate spinning external operators would allow for the probing of higher-spin exchanges in the cosmological collider. Furthermore, investigating the role of the Lorentzian inversion formula beyond the leading order in the expansion $1/N$ could refine the precision of these mass shift predictions. Ultimately, applying this concept to specific string-motivated models of inflation may yield quantitative, testable predictions for upcoming observations of the Cosmic Microwave Background and stochastic gravitational wave backgrounds, translating the algebraic consistency of the OPE into measurable imprints on the sky.


\end{document}